\documentclass[aps,prd,twocolumn,groupedaddress,showpacs,showkeys,floatfix,nofootinbib]{revtex4-2}

\usepackage{graphicx}
\usepackage{amssymb}
\usepackage{amsmath}
\usepackage{bm}
\usepackage{physics}

\newcommand{\afm}{a_{f,m}}
\newcommand{\afe}{a_{f,e}}
\newcommand{\axm}{a_{\chi, m}}
\newcommand{\axe}{a_{\chi, e}}
\newcommand{\yesnum}{\addtocounter{equation}{1}\tag{\theequation}}

\begin{document}

% Use the \preprint command to place your local institutional report
% number in the upper righthand corner of the title page in preprint mode.
% Multiple \preprint commands are allowed.
% Use the 'preprintnumbers' class option to override journal defaults
% to display numbers if necessary
%\preprint{}

%Title of paper
\title{Antisymmetric tensor portals to dark matter}

% repeat the \author .. \affiliation  etc. as needed
% \email, \thanks, \homepage, \altaffiliation all apply to the current
% author. Explanatory text should go in the []'s, actual e-mail
% address or url should go in the {}'s for \email and \homepage.
% Please use the appropriate macro for each type of information

% \affiliation command applies to all authors since the last
% \affiliation command. The \affiliation command should follow the
% other information
% \affiliation can be followed by \email, \homepage, \thanks as well.

%%\author{Student 1}
%%\email{student.1@someuni.somecountry}
%%\affiliation{Department of Physics and Engineering Physics, 
%%University of Saskatchewan, 116 Science Place, Saskatoon, Canada SK S7N 5E2}
%%\affiliation{Department of Physics, SomeUniversity, SomeCountry}

\author{Alexander J.~Magnus}
%\email[]{rainer.dick@usask.ca}
\email{alexander.magnus@usask.ca}

\author{Joshua G.~Fenwick}
%\email[]{rainer.dick@usask.ca}
\email{joshua.fenwick@usask.ca}

\author{Rainer Dick}
%\email[]{rainer.dick@usask.ca}
\email{rainer.dick@usask.ca}
%\homepage[]{Your web page}
%\thanks{}
%\altaffiliation{}
\affiliation{Department of Physics and Engineering Physics, 
University of Saskatchewan, 116 Science Place, Saskatoon, Canada SK, S7N 5E2}

%Collaboration name if desired (requires use of superscriptaddress
%option in \documentclass). \noaffiliation is required (may also be
%used with the \author command).
%\collaboration can be followed by \email, \homepage, \thanks as well.
%\collaboration{}
%\noaffiliation

%\date{\today}

\begin{abstract}
  Both freeze-in of very weakly coupled dark matter and freeze-out of initially
  thermalized dark matter from the primordial heat bath provide interesting
  possibilities for dark matter creation in the early universe.
  Both scenarios allow for a calculation of baryon-dark matter coupling
  constants as a function of dark matter mass $m_\chi$, $g=g(m_\chi)$,
  due to the constraint that
  freeze-in or freeze-out produce the observed dark matter abundance.
  Here we compare the resulting coupling constants in the two scenarios
  if dark matter couples to baryons through an antisymmetric tensor portal.
  The freeze-in scenario predicts much smaller coupling in agreement with
  the nonthermalization postulate. We find that the couplings as a function
  of mass behave very differently in the two scenarios.
\end{abstract}

% insert suggested PACS numbers in braces on next line %%separate by commas
%\pacs{ , }  
% insert suggested keywords - APS authors don't need to do this%%separate by commas


%\maketitle must follow title, authors, abstract, \pacs, and \keywords
\maketitle

% body of paper here - Use proper section commands
% References should be done using the \cite, \ref, and \label commands
%\section{Introduction}
% Put \label in argument of \section for cross-referencing
%\section{\label{}}
%\subsection{}
%\subsubsection{}

%%%%%%%%%%%%%%%%%%%%%%%%%%%%%%%%%%%%%%%%%%%%%%%%%%%%%%%%%%%%%%%%%%%%%%%%%%%%%%%%
%%%%%%%%%%%%%%%%%%%%%%%%%%%%%%%%%%%%%%%%%%%%%%%%%%%%%%%%%%%%%%%%%%%%%%%%%%%%%%%%

%%to do:
%%string signatures in intro, antoniadis
%%Include ref. to recent JHEP paper
%%maybe more refs. for freeze-in

\section{Introduction\label{sec:intro}}

The development of consistent models for dark matter creation in the early universe
is a triumph of astroparticle
physics that draws on expertise in every area of modern theoretical physics, including
statistical mechanics, cosmology, particle physics, and quantum field theory. 

A widely discussed mechanism, informed by early calculations of relic neutrino abundances,
is based
on 
dark matter freeze-out \cite{KT}. This framework assumes that dark matter was reciprocally
created and annihilated in thermal
equilibrium with the primordial heat bath after the reheating stage of the universe. This
process continued until the expansion
rate of the universe suppressed the reactions which maintained thermal equilibrium between
the baryonic and the dark sectors. The
decoupled nonrelativistic dark matter density then grew with respect to the still relativistic
heat bath with the scale factor $a(t)$
according to $\varrho_{CDM}(t)/\varrho_b(t)\propto a(t)$ because the relativistic baryons were
still performing expansion work due to their higher pressure.

Freeze-out theory is an integral part of the ``WIMP miracle'', \textit{viz.} that weakly
coupled massive particles (WIMPs) in the GeV
to TeV mass range could easily explain the observed dark matter abundance.
Though the present lack of detection of
nucleon recoils from dark matter places constraints on the preferred parameter range for WIMPs from
thermal freeze-out, it does not
completely rule out thermal freeze-out scenarios.   Another promising framework for the creation
of dark matter in the early universe
concerns the ``freeze-in'' mechanism, which occurs through annihilation of baryons in the thermal
heat bath \footnote{Note that prior
  to around 2010, the terms ``freeze-in'' and ``freeze-out'' were used interchangeably to describe
  what is presented as freeze-out in this paper.}.
Contrary to dark matter that is frozen-out from the primordial heat bath, freeze-in assumes that
the dark matter particles are so weakly coupled to
the baryonic sector that thermal equilibrium was never achieved between the baryonic and dark
sectors \cite{mcdonal2002,jmr2010}. Instead, baryon
annihilation since reheating gradually built up the dark matter density to its present value.
Particles with such extraordinarily weak couplings
are often referred to as ``feebly interacting massive particles'' or FIMPs. 

Both of the aforementioned dark matter creation mechanisms require the use of rate equations
involving baryon-to-dark matter annihilation cross sections.
However, they contrast in that freeze-out employs a thermal decoupling criterion and subsequent
evolution of the decoupled dark component during the
expansion of the universe to calculate the relic dark matter density, whereas freeze-in integrates
baryon annihilation to dark matter from the time of
reheating.  Accordingly, while using the same kind of fundamental physics in terms of baryon--dark
matter annihilation cross sections, the actual
calculations of the relic dark matter abundance are very different. Nonetheless, the different
central premises on dark matter thermalization imply
that, within every given particle physics model for dark matter, its coupling to baryons
as a function of dark matter mass should be greater in magnitude
in the freeze-out scenario than in freeze-in.
The primary focus of this study is to compare the dark matter coupling, $g(m_\chi)$,
predicted from
thermal freeze-out to those from freeze-in for models where a dark Dirac fermion $\chi$,
couples to baryons through the exchange of an antisymmetric tensor field, $C_{\mu\nu}$.

%%next: interest in antisymmetric tensor portal
We are particularly interested in an antisymmetric tensor portal because the discovery of fundamental
fields of this type in particle physics would indicate an extension of the Standard Model
through supergravity or string theory.
The proposals of supergravity and string theory were motivated as approaches to a theory of
quantum gravity, and it is an intriguing
question whether dark matter can shed light on %quantum gravity.
this open problem. In this regard, it is also of interest that Manton and Alexander recently pointed out
that parity violating gravitational interactions of antisymmetric tensor fields can provide an alternative 
approach to antisymmetric tensor detection through gravitational wave signals \cite{MA}.

An antisymmetric tensor field, $C_{\mu\nu}$, can couple to Standard Model fermions, $f$, through dipole terms,
\begin{equation} \label{eq:LfC}
\mathcal{L}_{fC}=-\sum_f g_f\overline{\psi}_f S^{\mu\nu}(\afm+\mathrm{i}\afe\gamma_5)
\psi_f C_{\mu\nu},
\end{equation}
where we use the spinor representation of the Lorentz generators,
\begin{equation}
S^{\mu\nu}=\frac{1}{2}\sigma^{\mu\nu}=\frac{\mathrm{i}}{4}[\gamma^\mu,\gamma^\nu],
\end{equation}

in the dipole operators. The parametrization of the couplings in (\ref{eq:LfC}) is redundant
in that we can absorb $g_f$ into the
magnetic dipole coupling $\afm$ and the electric dipole coupling $\afe$. We keep $g_f$
separate for now to point out that the
couplings (\ref{eq:LfC}) can arise from the low-energy limit of $SU(2)\times U_Y(1)$
invariant couplings in the form $g_f=v_h/M_f$,
where $v_h$ is the Higgs expectation value and $M_f$ is a coupling scale for the fermion $f$ \cite{strd}.
String origins of these
couplings from antisymmetric tensor oscillations of closed strings, or from the Kalb-Ramond 2-form
gauge fields that couple to string world sheets, are discussed in Refs.~\cite{adrd,rdick}.

The couplings (\ref{eq:LfC}) imply decay of the antisymmetric tensors with a decay constant
\begin{align*}
  \Gamma_C = &\sum_f %\qty
        [\afm^2%\qty
          (m_C^2-4m_f^2)+\afe^2(m_C^2+8m_f^2)]\\
    &\times \frac{g_f^2}{192\pi m_C^2}\sqrt{m_C^2-4m_f^2} \yesnum \label{eq:Cdecay}
\end{align*}
where the sum runs over fermions with masses $m_f<~m_C/2$. This implies that primordial
antisymmetric tensor fields that couple
to electromagnetic dipole moments are constrained by the requirement of longevity, $\tau>10^{18}$ s.
This limits the masses of
relic primordial tensors to very small values unless we assume extremely weak coupling to electrons
and neutrinos. On the other
hand, extremely weak couplings cannot be excluded, and the constraint on the couplings is less
severe in freeze-in dark matter
production with low reheating temperature. Freeze-in  production of antisymmetric tensor dark matter
was recently discussed by
Capanelli \textit{et al.} \cite{evan}.
In the present paper, we evade the longevity constraint by pursuing an
alternative dark matter
connection of antisymmetric tensor fields, \textit{viz.}
through an antisymmetric tensor portal to fermionic dark matter.

We also note that the absence of corresponding resonances in collider experiments limits the 
antisymmetric tensor mass 
to $m_C>200$ GeV for hadrophobic antisymmetric tensors, and to $m_C>1$ TeV for antisymmetric tensors 
that couple to dipole moments of quarks.
The couplings (\ref{eq:LfC}) give rise to an antisymmetric tensor portal if the antisymmetric tensor
also couples to a
dark fermion $\chi$,
\begin{equation} \label{eq:LchiC}
\mathcal{L}_{\chi C}=-\,g_\chi\overline{\chi} S^{\mu\nu}(\axm+\mathrm{i}\axe\gamma_5)
\chi C_{\mu\nu}.
\end{equation}
Constraints from Bhabha scattering limit the coupling to electrons and the antisymmetric tensor
mass $m_C$
to $m_Cm_e/\sqrt{a_{em}^2+a_{ee}^2}>7.1\times 10^4\,\mathrm{GeV}^2$ or $g_e\sqrt{a_{em}^2+a_{ee}^2}<m_C/(290\,\mathrm{GeV})$ \cite{strd}.
However, our primary focus here is on the internal consistency of freeze-out versus freeze-in dark
matter production, as expressed
by the requirement of smaller coupling in the freeze-in scenario. We will compare antisymmetric
tensor portal couplings and
masses for freeze-in and freeze-out in Sec.~\ref{sec:results}.

\section{Antisymmetric tensor couplings for freeze-in and freeze-out
  dark matter production\label{sec:results}}

Here we absorb the coupling constants $g_f$ and $g_\chi$ in Eqs.~(\ref{eq:LfC}) and (\ref{eq:LchiC})
into the corresponding dipole
coupling constants $\afm$ and $\afe$, or $\axm$ and $\axe$, respectively.

The procedures for the calculation of relic dark matter abundances from thermal freeze-out or freeze-in
are well established and
documented in the literature. The required ingredient from the specific dark matter model concerns the
annihilation rates that
connect the baryons to the dark matter particles. In our case this yields annihilations of Standard
Model fermions $f$ into
dark fermions $\chi$ with the annihilation cross section (with $s>4m_f^2$) 
\begin{eqnarray}\nonumber  
v\sigma_{f\overline{f}\to\chi\overline{\chi}}(s)&=&\frac{1}{64\pi s^2}
\frac{\mathrm{Re}\sqrt{s(s-4m_\chi^2)}}{(s-m_C^2)^2+m_C^2\Gamma_C^2}
\\ \nonumber
&&\times
\bigg[\afm^2\axm^2(s-4m_f^2)(s-4m_\chi^2)
\\ \nonumber
&&
+\,\afm^2\axe^2(s-4m_f^2)(s+4m_\chi^2)
\\ \nonumber
&&
+\,\afe^2\axm^2(s+4m_f^2)(s-4m_\chi^2)
\\ \nonumber
&&+\,\afe^2\axe^2(s+4m_f^2)(s+4m_\chi^2)
\\ \nonumber
&&+\,32\afe^2\axe^2m_f^2 m_\chi^2
\\ \nonumber
&&
-\,\frac{1}{3}(\afm^2+\afe^2)(\axm^2+\axe^2)
\\ \label{eq:vsigmasftochi}
&&\times
(s-4m_f^2)(s-4m_\chi^2)\bigg].
\end{eqnarray}

The cross section for the reverse process  $\chi\overline{\chi}\to f\overline{f}$ 
(with $s>4m_\chi^2$) is nearly identical, except for the replacement $m_\chi \to m_f$ under the radical,
\begin{eqnarray}\nonumber  
v\sigma_{\chi\overline{\chi}\to f\overline{f}}(s)&=&\frac{1}{64\pi s^2}
\frac{\mathrm{Re}\sqrt{s(s-4m_f^2)}}{(s-m_C^2)^2+m_C^2\Gamma_C^2}
\\ \nonumber
&&\times
\bigg[\afm^2\axm^2(s-4m_f^2)(s-4m_\chi^2)
\\ \nonumber
&&
+\,\afm^2\axe^2(s-4m_f^2)(s+4m_\chi^2)
\\ \nonumber
&&
+\,\afe^2\axm^2(s+4m_f^2)(s-4m_\chi^2)
\\ \nonumber
&&+\,\afe^2\axe^2(s+4m_f^2)(s+4m_\chi^2)
\\ \nonumber
&&+\,32\afe^2\axe^2m_f^2 m_\chi^2
\\ \nonumber
&&
-\,\frac{1}{3}(\afm^2+\afe^2)(\axm^2+\axe^2)
\\ \label{eq:vsigmaschitof}
&&\times
(s-4m_f^2)(s-4m_\chi^2)\bigg].
\end{eqnarray}

The velocity-weighted dark matter annihilation
cross sections are thermally averaged following the procedure of Gondolo and
Gelmini \cite{GG},
\begin{eqnarray}\nonumber
  \langle v\sigma\rangle(T)&=&\frac{1}{8m_\chi^4 TK_2^2(m_\chi/T)}
  \\ \label{eq:thermalGG0}
  &&\times
\int_{4m_\chi^2}^\infty\!ds\,\sqrt{s}\left(s-4m_\chi^2\right)\sigma(s)K_1(\sqrt{s}/T),
\end{eqnarray}
where the annihilation cross sections $\sigma(s)$ need to be calculated and substituted
into the integral using the specific dark matter model under consideration.
Here these are the cross sections (\ref{eq:vsigmaschitof}), summed over all Standard
Model fermions.

In freeze-out, the
thermally averaged annihilation cross section determines the evolution of the relic dark matter
density after freeze-out,
whereas in freeze-in we have to use the thermally averaged production cross section to evaluate
dark matter production from
the baryonic heat bath since reheating.
We noticed already that low reheating temperature alleviates the longevity constraint
on antisymmetric tensor dark matter \cite{evan}. However, since we employ antisymmetric tensor
 fields only as messengers between the dark and the baryonic sectors, we use a
 reheating temperature $T_R=10^{15}$ GeV, as it would generically appear in Susy-GUT models
 and in string theory if threshold effects are taken into account \cite{kaplunovsky}.

The requirement that the relic dark matter density evolves to the observed dark matter abundance
is a primary constraint on
dark matter model parameters. For the antisymmetric tensor portals (\ref{eq:LfC},\ref{eq:LchiC}),
this corresponds to a
52-dimensional parameter space through up to 48 couplings $a_{f,m}$ and $a_{f,e}$, 2 couplings $a_{\chi,m}$
and $a_{\chi,e}$,
and two masses $m_C$ and $m_\chi$. However, we are primarily interested in the impact of the different
dark matter creation
scenarios on the parameters. As such, we restrict the parameter space through the assumption of
universality of the magnetic
dipole couplings, $a_{f,m}=a_{\chi,m}\equiv a_m$, with $a_{f,e}=a_{\chi,e}=0$, or of the electric dipole
couplings, $a_{f,e}=a_{\chi,e}\equiv a_e$, with $a_{f,m}=a_{\chi,m}=0$.
We also require perturbativity, $a_{m,e}^2<8\pi\approx5.01$.

We find that the resulting three-dimensional parameter spaces can still yield the observed dark
matter density $\varrho_c=\Omega_c\varrho_{\mathrm{crit}}$
both through freeze-in or through freeze-out. We use the values from the
Particle Data Group, $\Omega_c=0.12h^{-2}$
and $\varrho_{\mathrm{crit}}=1.053672\times 10^{-5} h^2\,\mathrm{GeV}/\mathrm{cm}^3$, where
the reduced Hubble constant $h$ cancels in the evaluation of $\varrho_c$ \cite{PDG}.
We also use the Standard Model masses from the Particle Data Group for the calculation of the
cross sections.

\begin{figure}[htb]
\scalebox{0.4}{\includegraphics{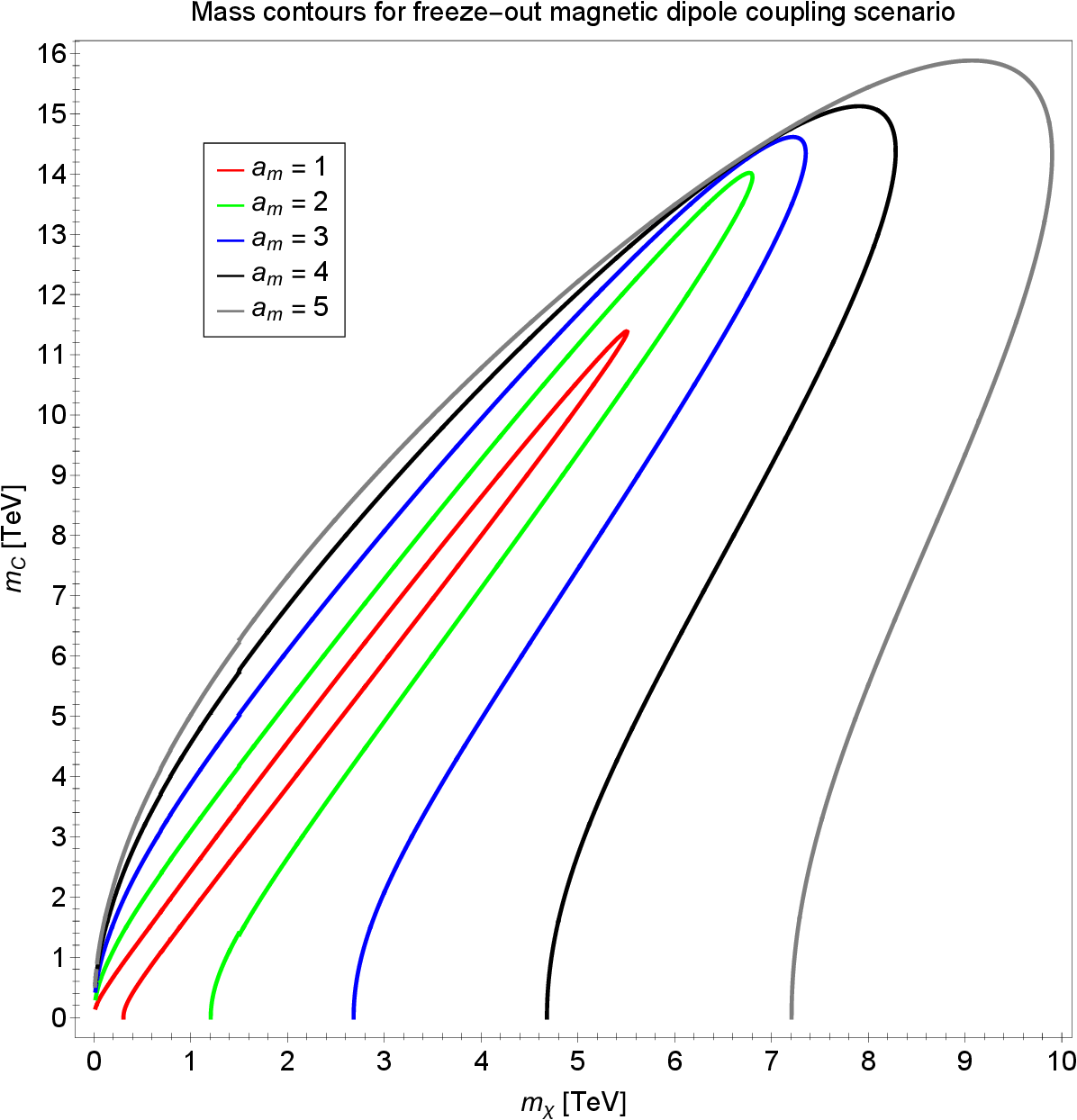}}
\caption{\label{fig:pam}
Relation between antisymmetric tensor mass $m_C$ and freeze-out dark matter mass $m_\chi$ for increasing $a_{m}$.}
\end{figure}

\begin{figure}[htb]
\scalebox{0.4}{\includegraphics{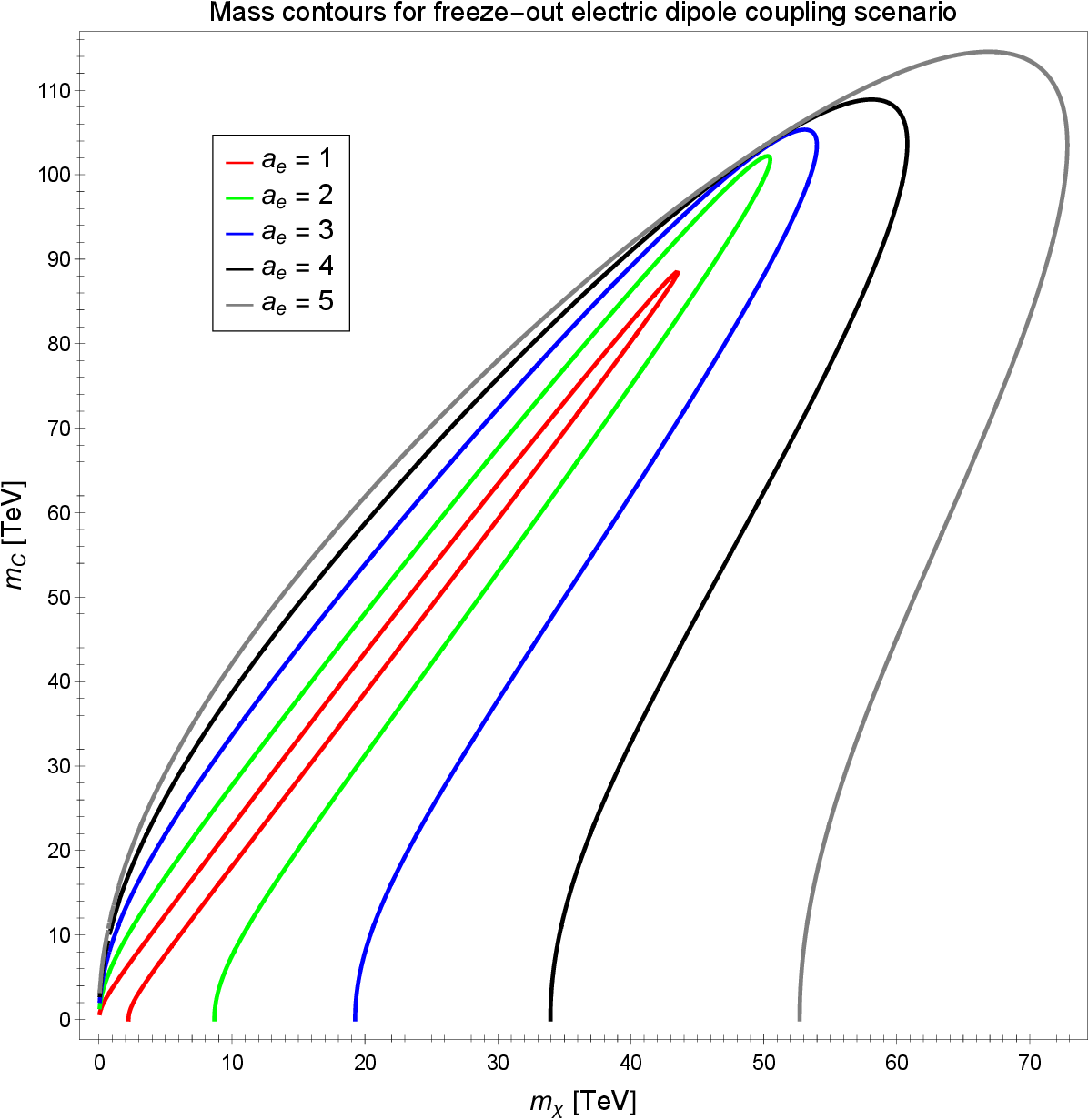}}
\caption{\label{fig:pae}
Relation between antisymmetric tensor mass $m_C$ and freeze-out dark matter mass $m_\chi$ for increasing $a_{e}$.}
\end{figure}

\begin{figure}[htb]
\scalebox{0.4}{\includegraphics{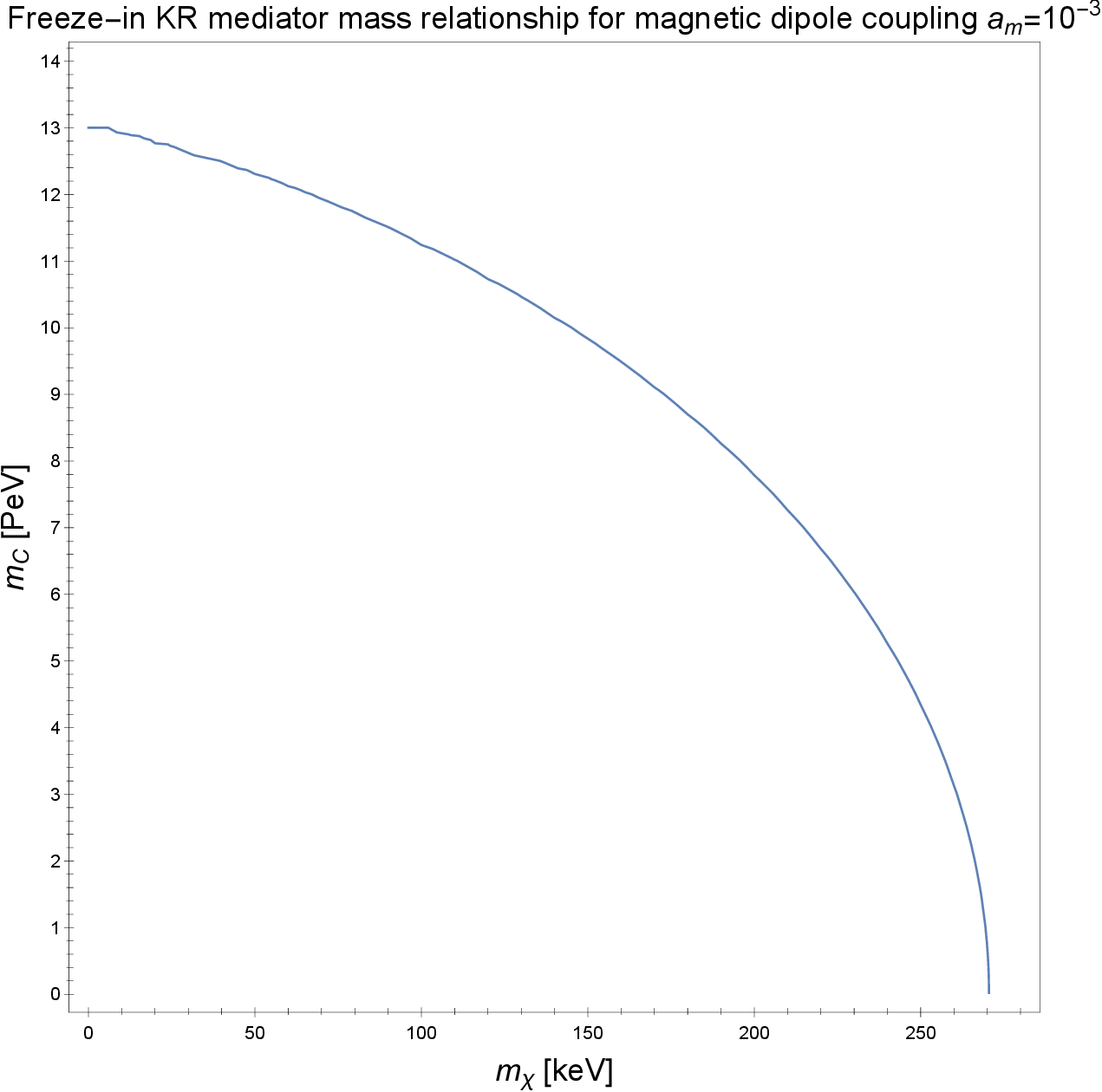}}
\caption{\label{fig:FIam3}
Relation between antisymmetric tensor mass $m_C$ and freeze-in dark matter mass $m_\chi$ for $a_{m}=10^{-3}$.}
\end{figure}

\begin{figure}[htb]
\scalebox{0.4}{\includegraphics{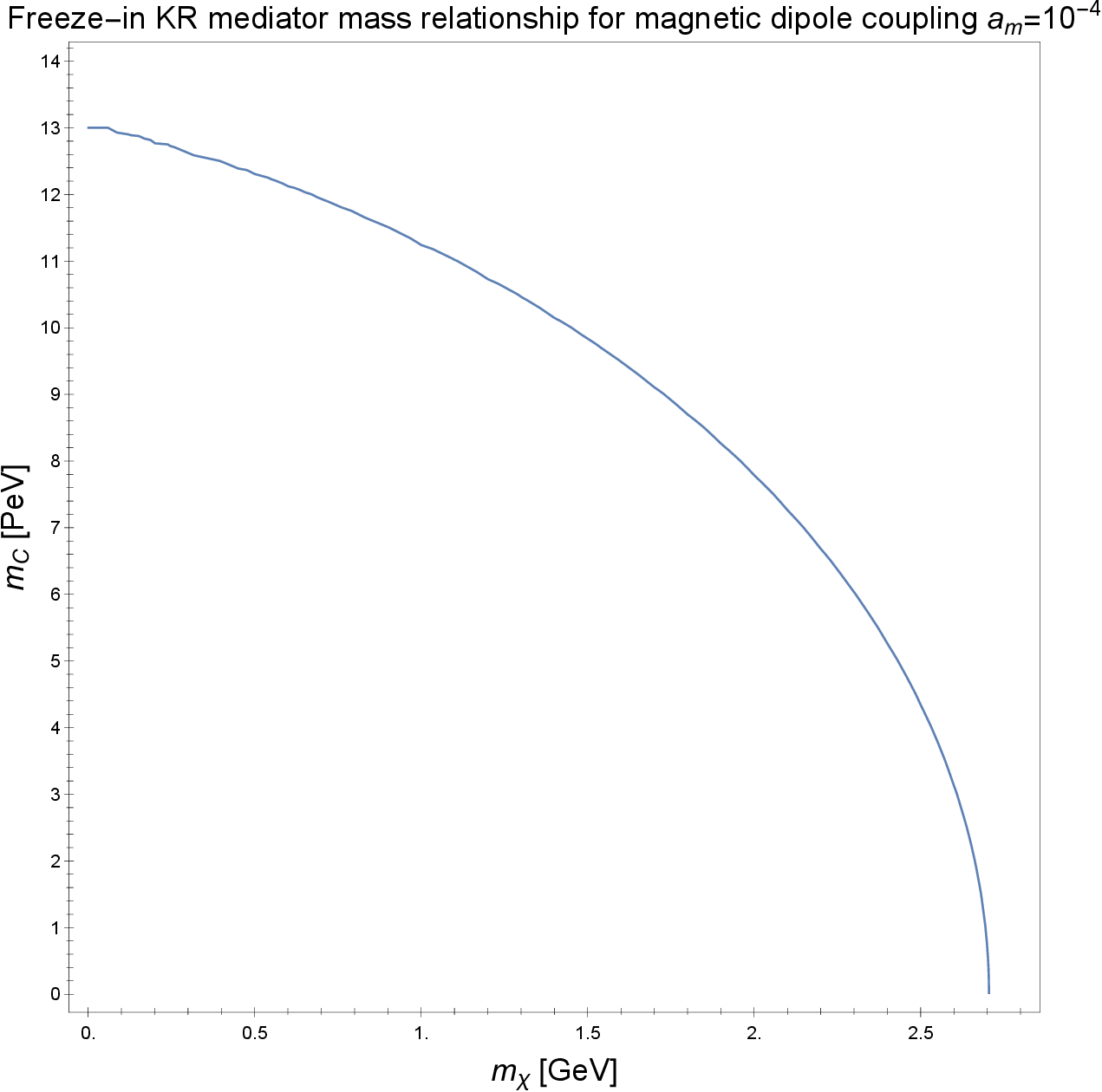}}
\caption{\label{fig:FIam4}
Relation between antisymmetric tensor mass $m_C$ and freeze-in dark matter mass $m_\chi$ for $a_{m}=10^{-4}$.}
\end{figure}

\begin{figure}[htb]
\scalebox{0.4}{\includegraphics{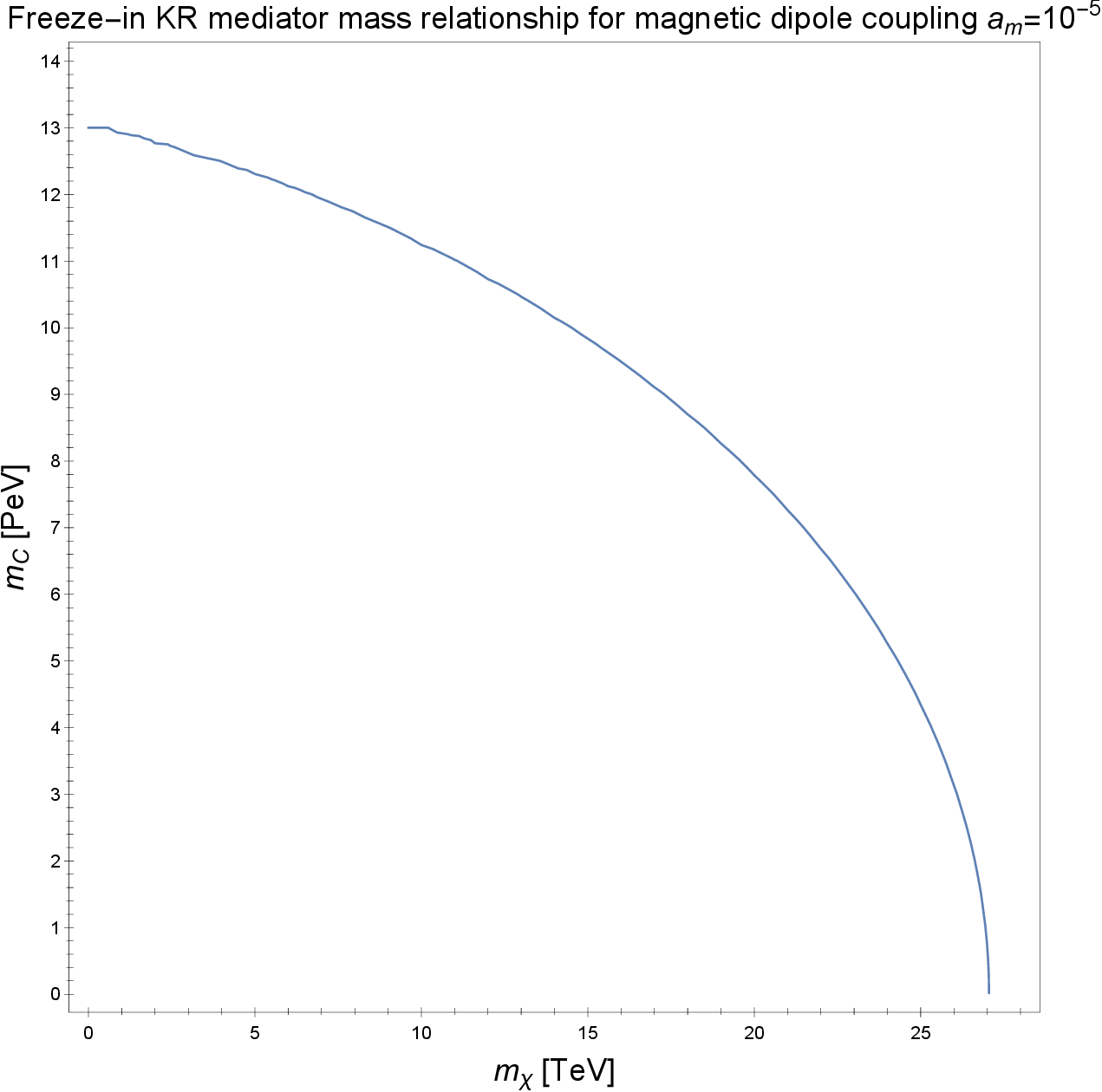}}
\caption{\label{fig:FIam5}
Relation between antisymmetric tensor mass $m_C$ and freeze-in dark matter mass $m_\chi$ for $a_{m}=10^{-5}$.}
\end{figure}

\begin{figure}[htb]
\scalebox{0.38}{\includegraphics{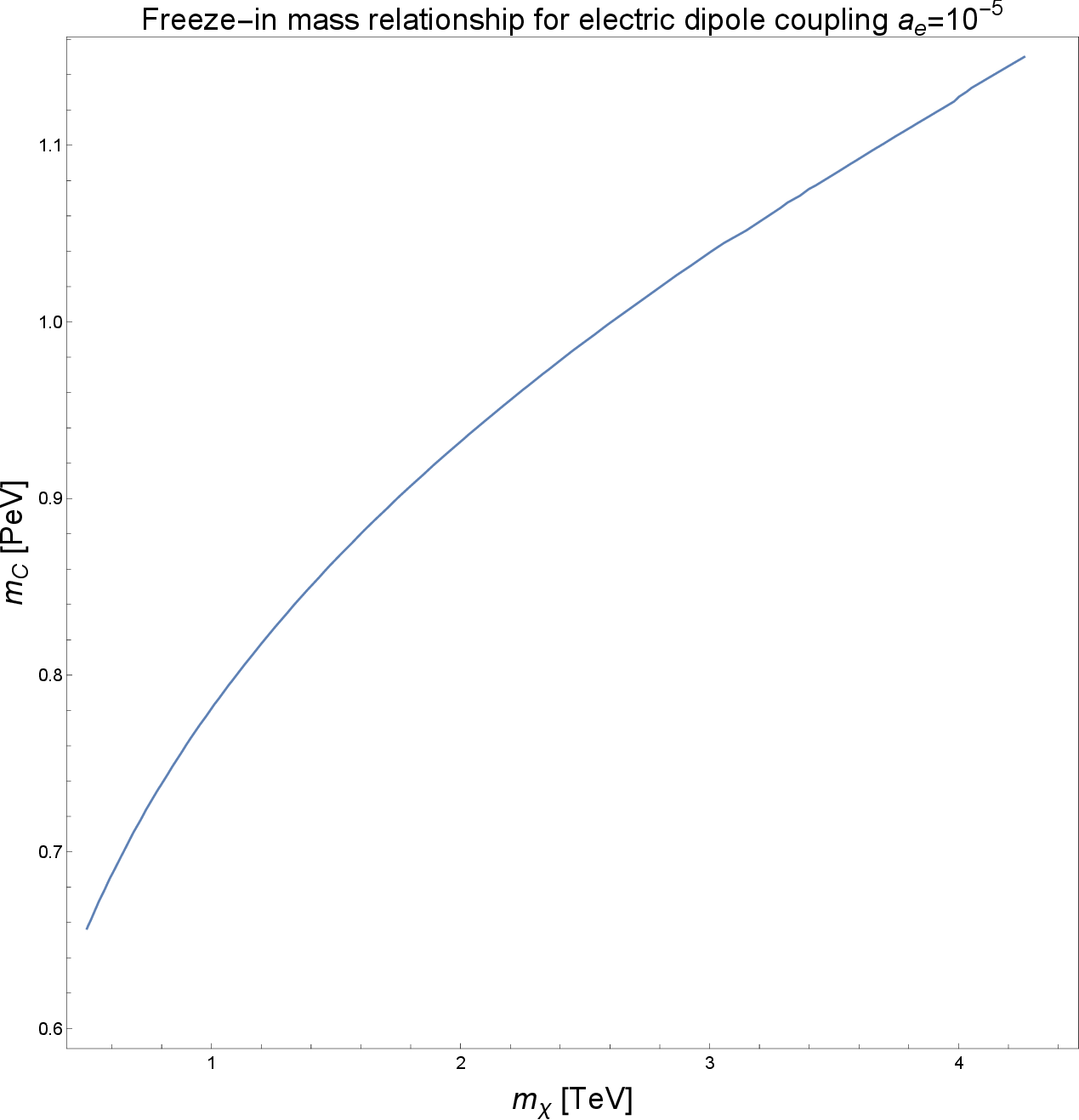}}
\caption{\label{fig:FIae5}
Relation between antisymmetric tensor mass $m_C$ and freeze-in dark matter mass $m_\chi$ for $a_{e}=10^{-5}$.}
\end{figure}

Mass relations $m_C=m_C(m_\chi)$ for different magnetic dipole couplings $a_m$ are 
displayed for freeze-out in Fig.~\ref{fig:pam}, and for freeze-in 
in Figs.~\ref{fig:FIam3}-\ref{fig:FIam5}.
Mass relations $m_C=m_C(m_\chi)$ for different electric dipole couplings $a_e$ are 
displayed for freeze-out in Fig.~\ref{fig:pae}, and for freeze-in 
in Fig.~\ref{fig:FIae5}.

We first note that freeze-in (Figs.~\ref{fig:FIam3}-\ref{fig:FIae5})
produces the required dark matter density for much weaker couplings than freeze-out 
(Figs.~\ref{fig:pam},\ref{fig:pae}), 
consistent with the basic freeze-in tenet that the coupling was too weak for dark matter
thermalization after reheating.

There are also several other interesting observations that follow from comparison of the dark
matter parameters in the two scenarios:

Freeze-in (Figs.~\ref{fig:FIam3}-\ref{fig:FIae5}) yields antisymmetric tensor masses in the PeV range 
and can generate dark matter over a very wide mass range. 
We found that every decrease of $a_m$ by an order of magnitude yields an increase of the dark matter mass $m_\chi$
by four orders of magnitude. Larger coupling produces more dark matter particles from freeze-in,
and this naturally
restricts the maximal possible dark matter mass from the abundance
requirement $(n_\chi+n_{\overline{\chi}})m_\chi=2n_\chi m_\chi=\rho_{CDM}$.
Since production rates scale with $a_m^4$ (because the impact of the decay constant $\Gamma_C$
is small for very weak couplings),
the increase in $m_\chi\propto a_m^{-4}$ for decreasing $a_m$ and constant $m_C$ 
is expected. Furthermore, since scaling $a_m\to a_m/10$ approximately scales $m_\chi(m_C)\to 10^4 m_\chi(m_C)$,
the ordinate $m_C\simeq 13$ PeV for $m_\chi\to 0$ is preserved.  

Contrary to the very small dipole couplings required for freeze-in, 
we found viable mass ranges for freeze-out only for relatively large dipole coupling near the 
perturbativity limit, $a_{m,e}^2\lesssim 8\pi$. This  yields TeV-scale masses both for the 
antisymmetric tensor and the dark matter in the case of magnetic dipole coupling,
see Fig. ~\ref{fig:pam}, and in the tens of TeV scale for electric dipole coupling, 
see Fig. ~\ref{fig:pae}.

We did not cut off the graph for freeze-out through a magnetic dipole coupling (Fig.~\ref{fig:pam})
below $m_C=1$ TeV, but note that these low antisymmetric tensor masses are ruled out in the present
model through absence of resonances at the LHC. 

As a novel feature of dark matter freeze-out through dipole couplings,
we find two possible dark matter mass values $m_\chi$ for every allowed $m_C$ both for the
magnetic dipole coupling (Fig.~\ref{fig:pam})
and the electric dipole coupling (Fig.~\ref{fig:pae}), except at the maxima of the $(m_\chi,m_C)$
curves. 
This is in contrast to freeze-out in simpler models, e.g.~minimal Higgs portal models
(see e.g.~\cite{mhp1,mhp2,mhp3,mhp4,mhp5,mhp6,mhp7,mhp8,mhp9,mhp10}),
where the dark matter coupling to baryons depends on a single coupling constant.
An increase in dark matter mass $m_\chi$ implies a decrease in the
required relic dark particle density $n_\chi$.
This implies later freeze-out from the primordial
heat bath, which requires larger cross sections \cite{KT}, 
which in turn can be achieved through larger coupling $g$
to baryons. Generically, this yields
a monotonically increasing coupling function $g(m_\chi)$
in freeze-out models. However, thermal averaging throws a
 wrench into this reasoning. Thermal averaging introduces a factor 
 \begin{equation} 
 \frac{1}{K_2^2(x_f)}\simeq\frac{2x_f}{\pi}\exp(2x_f)
\end{equation}
 in $\langle v\sigma\rangle(T_f)$, where $x_f=m_\chi/T_f\gg 1$ is the ratio
 between dark matter mass and freeze-out temperature, see Eq.~(\ref{eq:thermalGG0}).
 This ratio increases
 logarithmically with dark matter mass. The factor $1/K_2^2(x_f)$ can therefore
 increase $\langle v\sigma\rangle(T_f)$ with increasing dark matter
 mass, and this can generate the required increase in $\langle v\sigma\rangle(T_f)$
 without changing
the coupling parameters $m_C$ and $a_m$ or $a_e$, respectively.

For the comparison between freeze-out and freeze-in, we also note that both the much larger dipole 
couplings and the much smaller $m_C$ values in freeze-out versus freeze-in enhance the coupling of 
dark matter to baryons. This complies with the assumption of thermalization of dark matter
in the early universe before freeze-out, as opposed to the
absence of thermalization in the freeze-in scenario.

\section{Conclusions\label{sec:conc}}

We find that freeze-in through an antisymmetric tensor portal requires much smaller dipole couplings $a_{m,e}$ 
and larger antisymmetric tensor mass $m_C$ than freeze-out through the antisymmetric tensor portal,
in agreement with the assumption that interactions
with the primordial baryonic heat bath were too small to ever thermalize the dark matter. Freeze-in
requires an antisymmetric tensor mass in the PeV range,
whereas the dark matter mass $m_\chi$ can vary over a large range depending on the dipole coupling.
In particular, we find that that $m_\chi$ can range from
keV to TeV values for $10^{-3}\ge a_m\ge 10^{-5}$. For freeze-out, we find solutions with both $m_C$
and $m_\chi$ in the TeV mass range for magnetic dipole coupling,
and in the tens-of-TeV mass range for electric dipole coupling, if the corresponding 
dipole coupling is near the upper limit of the perturbative range. In the mass range for freeze-out,
we generically find two possible values for the dark
matter mass $m_\chi$ for given $a_{m,e}$ and $m_C$, although higher dark matter mass requires
later freeze-out corresponding to stronger coupling. However,
we point out that the calculation of the thermal average $\langle\sigma v\rangle(T_f)$
generates a positive feedback between dark matter mass and coupling
strength to baryons (as expressed 
through $\langle\sigma v\rangle(T_f)$), and this explains why the same set $(a_{m,e},m_C)$ of
coupling parameters to baryons can comply with more than one
possible dark matter mass for freeze-out.

Unfortunately, the large mediator masses, and in the case of freeze-in also the small dipole
coupling constants, push our estimates for recoil cross sections well below the neutrino floor.
However, the TeV mass scales of both the antisymmetric tensor and the dark fermion make the freeze-out
scenario interesting for next generation hadron
colliders.

\acknowledgments
We acknowledge support from the Arthur B.~McDonald Canadian Astroparticle Physics Research Institute and the Canada First Research Excellence Fund, and from the Natural Sciences and Engineering Research Council of Canada.\\

%\noindent


\begin{thebibliography}{88}

\bibitem{KT}
  E.~W.~Kolb and M.~S.~Turner, \textit{The Early Universe},
  (Addison-Wesley, Reading, 1990).

\bibitem{mcdonal2002}
  J.~McDonald, Phys.~Rev.~Lett.~\textbf{88}, 091304 (2002).

\bibitem{jmr2010}
  L.~J.~Hall, K.~Jedamzik, J.~March-Russell and S.~M.~West,
  JHEP \textbf{03}, 080 (2010).

\bibitem{MA}
  T.~Manton and S.~Alexander, Phys.~Rev.~D \textbf{110}, 044067 (2024).

\bibitem{strd}
  S.~Tiwary and R.~Dick, Eur.~Phys.~J.~C \textbf{81}, 1115 (2021).

\bibitem{adrd}
  A.~Dashko and R.~Dick, Eur.~Phys.~J.~C \textbf{79}, 312 (2019).
  
\bibitem{rdick}
  R.~Dick, Eur.~Phys.~J.~C \textbf{80}, 525 (2020).

\bibitem{evan}
  C.~Capanelli, L.~Jenks, E.~W.~Kolb and E.~McDonough, JHEP \textbf{2024(06)}, 075 (2024).

\bibitem{GG}
  P.~Gondolo and G.~Gelmini, Nucl.~Phys.~B \textbf{360}, 145 (1991).

\bibitem{kaplunovsky}
  V.~S.~Kaplunovsky, Nucl.~Phys.~B \textbf{307}, 145 (1988).

\bibitem{PDG}
  S.~Navas et al.~(Particle Data Group), Phys.~Rev.~D 110, 030001 (2024).

\bibitem{mhp1}
  V.~Silveira and A.~Zee, Phys.~Lett.~B \textbf{161}, 136 (1985).

\bibitem{mhp2}
  J.~McDonald, Phys.~Rev.~D \textbf{50}, 3637 (1994).
  
\bibitem{mhp3}
  M.~C.~Bento, O.~Bertolami, R.~Rosenfeld and L.~Teodoro, Phys.~Rev.~D \textbf{62},
  041302(R) (2000).
  
\bibitem{mhp4}
  C.~Burgess, M.~Pospelov and T.~ter Veldhuis, Nucl.~Phys.~B \textbf{619}, 709 (2001).
  
\bibitem{mhp5}
  A.~Djouadi, O.~Lebedev, Y.~Mambrini and J.~Quevillon, Phys.~Lett.~B \textbf{709}, 65 (2012).

\bibitem{mhp6}
  J.~M.~Cline, K.~Kainulainen, P.~Scott and C.~Weniger, Phys.~Rev.~D \textbf{88}, 055025 (2013);
  \textit{Erratum} Phys.~Rev.~D \textbf{92}, 039906 (2015).
  
\bibitem{mhp7}
  GAMBIT Collaboration (P.~Athron \textit{et al.}), Eur.~Phys.~J.~C \textbf{77}, 568 (2017).
  
\bibitem{mhp8}
  G.~Arcadi, M.~Dutra, P.~Ghosh, M.~Lindner, Y.~Mambrini, M.~Pierre, S.~Profumo
  and F.~S.~Queiroz, Eur.~Phys.~J.~C \textbf{78}, 203 (2018).
  
\bibitem{mhp9}
  R.~Dick, Int.~J.~Mod.~Phys.~D \textbf{27}, 1830008 (2018).

\bibitem{mhp10}
  A. Kvellestad, P. Scott and M. White, Prog. Nucl. Part. Phys. \textbf{113}, 103769 (2020).
  
\end{thebibliography}
\end{document}